\documentstyle[12pt]{article}
\input{tcilatex}

\begin{document}

\title{GRBs and SGRs as precessing gamma jets.}
\author{D. Fargion \\
%EndAName
Physics Dept. Rome, Univ. 1; INFN, Rome; Ple A. Moro 2.Italy\\
Physics Dept. Technion Institue,Haifa,Israel}
\date{}
\maketitle

\begin{abstract}
The GRB980425-SN1998bw association put in severe strain and contradiction
the simplest ``candle'' fireball model for GRBs. We probed that
statistically the association is reliable, the energy luminosity and
probability puzzles between cosmic and near by GRBs find a solution within a
precessing gamma jet model either for GRBs and SGRs. The expected
repetitivity for GRB980425 has been already probably observed on GRB980712.
\end{abstract}

\vspace{2cm} Preprint INFN \newline
1215, 13/07/98 \newline
Rome, Italy \newpage

\section{Introduction}

\subsection{The GRB luminosity/energy puzzle}

The recent GRB980425 event [2] has been observed in apparent peak gamma flux
comparable with previous exceptional GRB971214 [1] one: = $<l_{1 \gamma}>
\simeq 1.2 <l_{2 \gamma}> $ However these two GRBs locations are extremely
different: the GRB980425 event, if associated with nearest SN1998bw
explosion and its host galaxy ESO 184-G82, took place at near redshift $%
Z_2=0.0083$, while the far away host galaxy for GRB971214 burst is found at
redshift $Z_1 = 3.42$. Consequently their intrinsic average gamma luminosity
$< L_\gamma >$ and energies $< E_\gamma >$ ratio and the peak gamma
luminosity $L_\gamma$ (defined by the peak GRB flux in the sub-burst
events), following standard cosmological models, are huge (for isotropic
burst) because of their extremely different distances: \newline
\begin{equation}
\begin{array}{lll}
\frac{<L_{1 \gamma}>}{<L_{2 \gamma}>} & = & \frac{<l_{1 \gamma}>}{<l_{2
\gamma}>} \frac{[z_1 + (1-q_0^{-1})(\sqrt{2 q_0 z_1 +1} - 1)]^2} {[z_2 + (1-
q_0^{-1})(\sqrt{2 q_0 z_2 +1} - 1)]^2} \\
&  &  \\
& \cong & \frac{<l_{1 \gamma}>}{<l_{2 \gamma}>} \; \frac{<z_{1 }^2>}{<z_{2
}^2>} \cong 2 \cdot 10^5
\end{array}
\end{equation}
\newline
\begin{equation}  \label{eq2}
\frac{<E_{1 \gamma}>}{<E_{2 \gamma}>} = \frac{\int L_{1 \gamma} \, dt}{\int
L_{2 \gamma} \, dt} \cong \frac{\Delta \tau_1 <l_{1 \gamma}> z_1^2}{\Delta
\tau_2 <l_{2 \gamma}> z_2^2} \sim 4 \cdot 10^5
\end{equation}
\newline
\begin{equation}  \label{eq2b}
\left. \frac{L_{1 \gamma}}{L_{2 \gamma}} \right|_{peak} \simeq \left. \frac{%
l_{1 \gamma}}{l_{2 \gamma}} \right|_{peak} \; \frac{ z_1^2}{z_2^2} \simeq
10^7 \div 10^8
\end{equation}
\newline
where $\Delta \tau_1,\,\Delta \tau_2 $ are the observed GRBs durations. The
approximations hold because of the negligible dependence on z and on the
deceleration parameter q (for any deceleration values smaller than unity) in
the squared bracket in equation 1. Most observed cosmological parameters do
require $q \leq 1$.

\subsection{The credibility of GRB980425-SN1998bw association}

These five-six order of magnitude for near-far GRBs gamma average luminosity
ratio put serious doubt on the existence of any unique isotropic ''candle
GRB-model'', as the most celebrated fireball-hypernovae ones. Before
questioning the credibility of those models, let first inquire the
credibility of GRB980425-SN1998bw association. The last near GRB has been
found in only in the wide field (WF) Beppo-Sax camera contrary to better
localized optical transient (OT), all at optical association, within the SAX
narrow-field detector (NF): GRB970228 [3] ,GRB970508 [4],GRB971214 [1] and
recent GRB980703 [5]. In particular let as compare again the April 98 and
December 97 GRB event probability $P$, ratio $R\equiv P_{1}/P_{2}$ to find
any OT within each of them by chance at SAX narrow-field solid angle $%
A_{1}=\theta _{1}^{2}=(1^{\prime })^{2}$, at SAX wide field (WF) solid angle
$A_{2}=\theta _{2}^{2}=(8^{\prime })^{2}$and at their %%@
corresponding redshift $(z_{1}=3.42,z_{1}=0.0083)$ volumes: \newline
\begin{equation}
R_{1/2}=\frac{P_{1}}{P_{2}}\cong \frac{A_{1}z_{1}^{3}}{A_{2}z_{2}^{3}}
\simeq 1.1 \cdot 10^{6} \;\;\; ;
\end{equation}
\newline
this huge number implies that it is more reasonable and honest to wonder on
the ''cosmic'' associations than to negate the local GRB980425/SN1998bw one
(whose probability to occur by chance is smaller than $10^{-4}$). Even
restricting the cosmic GRBs area $A_1$ at tiny optical sizes (arc seconds)
one still finds a large probability ($>300$) ratio favoring the April 98
association respect to any cosmic one. In brief it is statistically
significant the GRB980425 location at SN1998bw at a distance within about 40
Mpc. The same cannot be said for the cosmic GRBs events.

\section{GRBs open questions and the crisis of the fireball model}

Once we agreed on the nature of nearest April 98 GRB event and taking for
grant the far cosmic GRB locations, we face, following "candle" models,
again the luminosity/energy puzzle in equations (1) and (2). Maybe we are
just observing two different kind of GRBs ? Maybe the nearest one (if
isotropic and "homogeneous" in all the wavelength emissions) is weaker than
the far event at z=3.42 ?At least the optical intensity state the opposite:
the GRB April was comparable with the far away one on December 97. Indeed
their average peak optical intensities (intrinsic $L_{OT}$, apparent $l_{OT}$%
) ratio \mbox{are:} \newline
\begin{equation}
\frac{<L_{OT_1}>}{<L_{OT_2}>} \; \cong \; \frac{<l_{OT_1}>}{<l_{OT_2}>} \;
\frac{z_1^2}{z_2^2} \simeq 10^{\frac{2}{5}\;(M_2 - M_1)} \frac{z_1^2}{z_2^2}
\simeq 10
\end{equation}
\newline
where $M_2 \simeq 12.5$, $M_1 \simeq 21.5$ are the observed peak magnitudes
in R band of the OTs; therefore the peak gamma luminosity (equation \ref
{eq2b}) and the optical ratios exhibit strident opposite behaviours for any
"candle" fireball source. Finally we notice that the integral optical
luminosity $<L_{OT}>$ for the two events is comparable while their gamma
fluxes are extreme (equation \ref{eq2}). Moreover the ratio between peak
gamma luminosity over OT one is: \newline
\begin{equation}
\frac{L_{1 \gamma}}{L_{2 \gamma}} : \frac{L_{OT_1}}{L_{OT_2}} \geq 10^7
\end{equation}
\newline
How can one concert so many order of magnitudes in an unique "candle"
isotropic GRB model? In astrophysics one may be allowed to play one-two
order of magnitudes, no more. Therefore "classical" standard fireball model
is no longer acceptable. However we notice, like for a Fenice, the arise, by
some authors of new "generations" of GRBs classes of "fireball scaled"
models [6]: S-GRBs for near SN GRB, C-GRB for Cosmic GRBs. These model
proliferation is analogous to what already occurred before for Soft Gamma
Ray Bursts. Originally these bursts whose nature was also bursting and
sudden like GRBs (as well as their overeddington luminosity), where all
within GRB dictionary. Once their local (galactic) and softer repeatability
nature was better defined, the theorists split their fate from all other
"mysterious" GRBs. The SGRs are not yet well understood as fireballs, even
new "isotropic" models (magnetar starquake) attempted to solve their origin.
We do believe SGRs are linked to GRBs. %Some authors []
%defined new cosmic (C-GRB) and supernova nearer GRB models
%(S-GRB). Within this philosophy one may easily predict a
%proliferation of future classes to encompass any new observation
%leading to a characteristic "label" to each GRB
%distance/morphology. The GRB understanding do not go far away with
%such an approach: for each GRB, one model is not a theory but just
%an empirical enumeration.
There may and indeed exist an unified GRB model able to answer most of the
puzzling questions above as well as the related following ones.

\subsection{The GRB signature puzzle}

Let us look deeper on GRBs puzzling nature of the near/far GRBs within an
isotropic candle model.

\begin{description}
\item[a)]  Why the nearer GRB was "softer" than the cosmic "harder" far
anyway ones ? The observed soft-hard nature of nearer April GRB and cosmic
far GRBs is in strident disagreement with cosmological expansion
predictions: any red shift for cosmic GRB fireball, with candle spectra,
would always appear softer (at least by a factor $\frac{\Delta \nu}{\nu}
\simeq (1+z_1) \sim 4.4 $) and not, as observed, harder than near ones.

\item[b)]  Why the time structure of near GRB was smoother than the rapid
structured cosmic GRB on December 1997 ? The cosmic expansion, once again,
would lead to an opposite behaviour: a standard red shifted GRBs must appear
relativistically Doppler-shifted and time diluted ($\frac{\Delta \tau}{\tau}
\simeq (1+z_1) \sim 4.4 $) while the GRB980425-GRB971214 data smoothly
structured show exactly the opposite.

\item[c)]  Why the intensity of GRB where in opposite ratio (as already %%@
noticed in Eq. 1-2) while a single candle model would just naturally imply
the opposite (larger gamma flux for nearer burst) by a factor $\frac{z_1^2}{%
z_2^2} \sim 1.7 \cdot 10^5$?

\item[d)]  Why statistically, we were able to observe so "many" (even just
one) near GRBs in such a nearby small redshift volume ? A simple statistical
argument imply that if GRB are homogeneous in space and time the far GRB
should be much more numerous than few (or just unique) observed ones as
GRB980425. For our two extreme events: \newline
\begin{equation}  \label{eq6}
\frac{P_1}{P_2} \cong \frac{z_1^3}{z_2^3} \simeq 7 \cdot 10^7
\end{equation}
\newline

Some strong bias must suppress this huge number for far gamma GRBs.
Otherwise we were exceptionally lucky to see a nearest GRB980425. As God
does not like to play dice, neither GRBs do.

\item[e)]  Why GRBs are not (always) time coincident with the rise of the
optical transient ? This is well known and observed for GRB970508 (which
grew in optical intensity days after its GRB) as well as for the
extrapolation of optical-radio flux of GRB980425. A gamma fireball would not
wait so long to lead to an optical signal. Anyway within ``one shoot''
fireball model an unique trigger time for OT, X, $\gamma$ and radio events
would be naturally expected.

\item[f)]  Finally, if local GRBs are not a rare event, and if they are
isotropic, one must find a "growth" in the number count-flux diagram at
lowest flux above the inhomogeneous decay. Indeed the number of SGRs (near
GRBs) at redshift $z \geq 0.0083$, might pollute the lowest region of number
count test.\newline
This imply that the ''cosmological'' (or space-time) inhomogeneity in number
count may be hidden by a more significant nearby source population, contrary
to the naive cosmological interpretation %%@
of GRB counts. \newline
\end{description}

\section{Toward the answer: a beaming gamma jet}

The first simplest solution to solve the GRB luminosity puzzle within an
unified GRB model is to look for a ''geometrical'' enhancement (by a narrow
beam) able to lead, when observed at different angular sides, to large
intensity modulations. This beaming occur naturally for relativistic jet
with angle $\theta \sim 1 / \gamma$ originated by micro-quasars like objects
as those recently discovered in our galactic halo . Any ''large'' cone beam (%
$\theta \geq $few degree size), is not able to reconcile at least the six or
seven order of magnitudes in extragalactic GRBs intensities.\newline
A tiny highly collimated beam is necessary $(\theta \sim 10^{-3} \div
10^{-4})$ also within an inverse Compton Model for GRBs able to scatter low
energy photons (I.R. or BBR) to highest energies. Moreover if one desire to
correlate the GRB nature with their soft-gamma SGRs sources an intensity
decay must be required to scale the GRB power toward S-GRB with time.
\newline

\subsection{Are GRBs an episodic beamed pulse?}

Is GRB just an impulsive (unique burst) [7] beamed event? If this is the
case we may increase by many orders $(8\div10)$ its apparent luminosity but
we face a ''probability puzzle'' related to the rarity to observe (within a
cone $\frac{\theta^2}{4 \pi} \sim 10^{-7} \div 10^{-9})$ a SN burst jet at
low redshift (for the optical burst there is no need or indications for
beaming), pointing toward us.\newline
Moreover this ''burst'' solution will not explain the fine structured and
fast variable nature of some GRB neither the puzzling repeating nature of
SGRs. \newline

\section{GRBs and SGRs as multiprecessing Gamma Jet}

Therefore we are forced [8] to consider the GRBs as due to multiprecessing
Gamma Jets (as the recent discovered microquasar objects in our own galaxy).
In our first approach we believed that all GRBs were all like SGRs, i.e.
within a wide galactic halo. Until February 97 GRB we were afraid to
require, for a mini-jet power, too large energies rate comparable with SN
ones. We now consider their nature (following latest evidences for some
cosmic location) either at their oldest stages in our own galactic halo [9]
in the role of Soft Gamma Repeaters (SGRs) and, at their earliest epochs,
near their SN-like birthdate, while at their peak intensity. The most
powerful Gamma Jets beamed to us are observed far away (cosmic C-GRBs),
while nearer and local ones (S-GRBs like SN1998bw) are more rare for
statistical reasons (smaller volume). Repeaters are due to their nearer
location and consequent more intense apparent gamma flux, which may be seen
also at wider beam periphery.\newline
The Inverse Compton Scattering, probably fueled by high energy electron
pairs, converts low energy photons (infrared or cosmic black body ones) into
a coaxial collimated gamma jet at MeV energies. The relativistic kinematics
imply that the inner jet cone contains most intense and hardest radiation.
The photons at outer coaxial jet cones are less abundant and less energetic.
Nearest sources (galactic-local SGRs / S-GRBs) may be observed either
bursting, rarely, in their peak inner core or, often, blazing and/or flaring
from wider peripheral jet regions. Far away cosmic GRBs are observable (by
present detectors) only during their peak apparent intensity no longer than
their gamma jet birth, marked by optical SN explosion, within their inner
and harder beam jet. The energy decay of the jet output makes far away GRBs
observable within few days from SN optical transient, which explosion is a
nearly isotropic burst, and whose consequent radio tail is partially beamed
by synchrotron radiation.\newline
\vspace{0.5cm} Let us describe more in quantitative detail the GRB genesis
and evolution towards SGR regime. We first imagine a star collapse or,
better, a binary stellar system feeding a collapse and explosion. The
asymmetry of the system defines an axis of the relic compact object (a
neutron star NS or black hole BH) which becomes the source of a thin jet.
The exact acceleration and collimation of the jet is still a mystery;
ultrarelativistic beamed cosmic rays source (from the compact NS and BH) or
electromagnetic acceleration and confinement is needed. The recent
observational evidence of the reality of such microquasar jets in our Milky
Way as GRS 1758-258, GRS 1915+105 and GRB J1655-40 is well based and widely
accepted. The ejected matter contains an ultrarelativistic electron pair
beam which is highly collimated, $\theta \leq \frac{1}{\gamma _{e}}$; the
Inverse Compton Scattering of these electron pairs on thermal photons (I.R.
or cosmic BBR) is a source of a new collinear gamma jet (along the electron
pair one). As an order of magnitude we assumed the electron pair energy $%
E_{\gamma }\sim 10\,GeV$ and their target thermal photon just like the $%
2.75\,K$ Black Body Radiation; then $K_{B}T\simeq 2.75\,K$, $\gamma
_{e}\simeq \frac{E_{e}}{m_{e}\,c^{2}}=2\cdot 10^{4}$. It is possible to show
[9],[10] that the differential number distribution for gamma jet photons
from Inverse Compton Scattering of monochromatic ultrarelativistic electron
pair on isotropic BBR is:\newline
\begin{equation}
\frac{dN_{1}}{dt_{1}\,d\epsilon _{1}\,d\Omega _{1}}\simeq A_{1}\epsilon
_{1}\ln \left[ \frac{1-\exp \left( \frac{-\epsilon _{1}(1-\beta \cos \theta
_{1})}{k_{B}\,T\,(1-\beta )}\right) }{1-\exp \left( \frac{-\epsilon
_{1}(1-\beta \cos \theta _{1})}{k_{B}\,T\,(1+\beta )}\right) }\right] \left[
1+\left( \frac{\cos \theta _{1}-\beta }{1-\beta \cos \theta _{1}}\right) ^{2}%
\right]   \label{eq7}
\end{equation}
\newline
where $A_{1}$ is a normalization factor defined by the intrinsic electron
jet flux intensity, $\epsilon _{1}$ is the electron pair energies, $T$ is
the target thermal photons, $\theta _{1}$ is the angle between the electron
jet axis and the observer. After the energy integral $\epsilon _{1}$, the
adimensional differential number rate becomes [9] \newline
\begin{equation}
\frac{\left( \frac{dN_{1}}{dt_{1}\,d\theta _{1}}\right) _{\theta _{1}(t)}}{%
\left( \frac{dN_{1}}{dt_{1}\,d\theta _{1}}\right) _{\theta _{1}=0}}\,=\,%
\frac{1+\gamma ^{4}\,\theta _{1}^{4}(t)}{[1+\gamma ^{2}\,\theta
_{1}^{2}(t)]^{4}}\simeq \frac{1}{(\gamma \,\theta _{1})^{4}}  \label{eq8}
\end{equation}
\newline
where the value at fixed angle $\theta _{1}=0$ is the peak gamma flux and $%
\beta $, $\gamma $ are the ultrarelativistic velocity and Lorentz factor of
electron pairs. Consequently the adimensional photon number as a function of
the small precessing angle $\theta _{1}$ grows as \newline
\begin{equation}
\frac{\left( \frac{dN_{1}}{dt_{1}\,d\theta _{1}}\right) _{\theta _{1}(t)}}{%
\left( \frac{dN_{1}}{dt_{1}\,d\theta _{1}}\right) _{\theta _{1}=0}}\simeq
\frac{1+\gamma ^{4}\,\theta _{1}^{4}(t)}{[1+\gamma ^{2}\,\theta
_{1}^{2}(t)]^{4}}\,\theta _{1}\approx \frac{1}{(\theta _{1})^{3}}
\label{eq9}
\end{equation}
\newline
the last approximation holding for $\gamma \,\theta \gg 1$. This number
density rate is proportional to the observable gamma luminosity of GRBs
(peak luminosity in equation \ref{eq2b}). Finally the total photon gamma
fluence outside the beam cone at maximal impact angle $\theta _{1m}$
recorded from GRBs (due to such precessing gamma jets) is \newline
\begin{equation}
\frac{dN_{1}}{dt_{1}}(\theta _{1m})\simeq \int_{\theta _{1m}}^{\infty }\frac{%
1+\gamma ^{4}\,\theta _{1}^{4}}{[1+\gamma ^{2}\,\theta _{1}^{2}]^{4}}%
\,\theta _{1}\,d\theta _{1}\simeq \frac{1}{(\,\theta _{1m})^{2}}\;\;\;.
\label{eq10}
\end{equation}
\newline
In a first approximation this influence is proportional to the observed GRB
energy. \newline
Now let us assume for the gamma jet power an initial ''standard candle''
power of intensity $I_{1}$. We assume, for sake of simplicity, an initial
beam power comparable to the maximal optical power associated with the
isotropic SN/GRB jet birth:\newline
\begin{equation}
I_{1}\,\simeq \,10^{44}\;\;\;erg\;s^{-1}\;\;\;.  \label{eq11}
\end{equation}
\newline

\bigskip The above gamma beam power $I_{1}$ is proportional to  $A_{1}$ the
electron jet one in equation 8. Their proportionality is related to
ICS  efficency which here is assumed within unity.

A jet power like this, while beamed within a thin jet cone of angle $\theta
_{e}\simeq 10^{-4}\;rad$ may explain an apparent power as large as $P\sim
4\pi I_{1}\theta _{e}^{-2}\simeq 10^{52}\div 10^{53}\;\;erg\;s^{-1}$. Our
assumption in equation \ref{eq11} is based on a very reasonable energy
equipartition argument.\newline
Let us also assume a decay power law $\left( \frac{t}{t_{0}}\right)
^{-\alpha }$ for the jet intensity; a conservative one, inspired by optical
GRB evolution, implies $\alpha \geq 1$.\newline
We may calibrate the characteristic power time scale requiring that the
initial young GRB intensity at later stages will correspond, as an order of
magnitude, to the ones ($\sim \;1000$ years old) observed gamma precessing
jets which behaves, in our own galaxy, as SGRs. This power is nearly $%
10^{38}\,erg\,s^{-1}$. Therefore the jet power at any time t is:\newline
\begin{equation}
I_{jet}\,=\,I_{1}\;\left( \frac{t}{t_{0}}\right) ^{-\alpha }\simeq
10^{44}\left( \frac{t}{3\cdot 10^{4}\,s}\right) ^{-1}\;\;erg\,s^{-1}
\label{eq12}
\end{equation}
\newline
where $t_{0}$,in the last expression, is derived assuming a power exponent $%
\alpha \approx 1$. The beaming angle is assumed below $10^{-4}\div 10^{-3}$
radiant, depending on exact relativistic jet nature. The apparent GJ power
is $4\pi \gamma _{e}^{2}$ enhanced corresponding to peak luminosity $%
L_{\gamma }\simeq 10^{53}\div 10^{54}\;\;erg\;s^{-1}$.

\section{The Jet Beam probability to hit the observer}

The above model tools allow us to understand the puzzling low probability in
equation \ref{eq6} ($\sim (7 \cdot 10^{7})^{-1}$) to observe the near April
GRB respect to cosmic GRBs; we remind that, after one year of Beppo Sax era,
one found, within nearly 300 GRBs, only one such a GRB. Therefore one need
an amplification factor $A_2$, related to a wider observation area cone,
amplification able to complements the puzzling ratio in equation \ref{eq6}:%
\newline
\begin{equation}  \label{eq12a}
A_2 \cong \frac{7 \cdot 10^7}{300}\simeq 2.5 \cdot 10^5 \;\;\; .
\end{equation}
\newline
This geometrical solid angle amplification factor is naturally related to a
corresponding observation angle $\theta_1$ for April event respect to a very
narrow observation angle for cosmic thin beamed GRBs:\newline
\begin{equation}  \label{eq13}
a_2 \simeq \sqrt{A_2} \sim 500 \;\;\; .
\end{equation}
Assuming for a December GRB event an angle $\theta_{Dec} \simeq \frac{1}{%
\gamma_e} \sim 10^{-4}$, this implies that near April GRB event was seen
within \newline
\begin{equation}  \label{eq14}
\theta_{Apr} \simeq a_2 \theta_{Dec} \simeq 5 \cdot 10^{-2}
\end{equation}
a cone only few degrees wide. In the frame of one ``shoot'' GRB model we
would observe the GRB event with a low probability ($2 \cdot 10^{-4}$). On
the contrary within a continuous precessing gamma jet model one finds
additional probabilities due to the integral time. As a first approximation
the probability to be blazed and flashed by such a ''wide'' precessing beam
cone, assuming a characteristic time delay between the supernova event (and
its optical light beginning) as observed ($\Delta \tau \approx 2 \, day$)
and a characteristic GRB duration ($\Delta \tau_{GRB} \sim 20 \, s $), is
\newline
\begin{equation}  \label{eq15}
P \simeq \left( \frac{a\,\theta_{Dec}}{4\,\pi}\right)^2 \, \frac{\Delta \tau%
}{\Delta \tau_{GRB}} \simeq 1 \;\;\;.
\end{equation}
\newline
Within a two-day period ($\sim 4 \, t_0$) the gamma jet intensity decreased
and the final probability in the above approximation will be smaller but
still within unity. Therefore the precessing GJ on April 25 had the possibly
to blaze (as indeed happened) Earth within its wide light-house gamma beam.
Could such a powerful jet repeat the hit? We know that SGRs (nearest
sources) can. As we shall see in paragraph \ref{par8}, the GRB possibly
repeated on 12 July 1998.\newline

\section{The Gamma Jet intensity and its Repeater Nature}

Let us verify if the near/far GRBs intrinsic luminosities within the present
standard candle precessing gamma jet values are comparable to the observed
ones .For an amplification angle $a \simeq 500$ one would expect a total
number photon fluence ratio (between December cosmic event toward April near
event) to be compared with average energy ratio in equation \ref{eq2}
\newline
\begin{equation}  \label{eq16}
\frac{N_{\gamma \,1}}{N_{\gamma \,2}} \simeq \frac{<E_{\gamma \,1}>}{%
<E_{\gamma \,2}>} \simeq a^2 \simeq 2 \cdot 10^5
\end{equation}
\newline
while the ''peak''luminosity intensity as defined in equation \ref{eq2b}
(related to the structured nature of the GRB, i.e., to the intrinsic peak
luminosity at each internal GRBs mini-burst) is \newline
\begin{equation}  \label{eq17}
\frac{L_{\gamma \,1}}{L_{\gamma \,2}} \simeq a^3 \simeq 10^8 \;\;\; ,
\end{equation}
\newline
these values are well within the observed ones. Moreover the characteristic
time scales for near/far GRB signals must reflect (assuming a common
precessing angular velocity) their different impact angle parameters: $%
\theta_{Dec} \simeq \frac{1}{\gamma}$; $\theta_{Apr} \sim \frac{a}{\gamma}$.
Indeed the minimal observed time scales of December rapid structured event
is $\Delta \tau_{Dec} \simeq 10 \,m s$ while the April smooth event,
observed at a wider impact parameter is observed at time scales $\Delta
\tau_{Apr} \simeq 20 \, s$. Their ratio is, as an order of magnitude:\newline
\begin{equation}  \label{eq18}
\frac{\Delta \tau_{Dec}}{\Delta \tau_{Apr}} \approx \frac{1}{a} \;\;\; .
\end{equation}
These general features, while giving an answer to most puzzles, favor the
GRB interpretation as due to a precessing gamma jet in an unified model.%
\newline

\section{The probable repeating\newline
nature of GRB980425: GRB980712}

\label{par8} Because the higher probability to observe again a near
(intense) Gamma Jet, we may wait for a second GRB flash from GRB980425.
Indeed the probability $P$ to re-observe the GRB within BATSE sensitivity is
\newline
\begin{equation}  \label{eq19}
P \simeq \int N \theta^2 \,dt \simeq \int \theta^{-2} I_0 t^{-\alpha}
\theta^2\,dt \approx t^{-\alpha + 1}
\end{equation}
\newline
where the intrinsic gamma luminosities and flux are $I \sim I_0 \,
t^{-\alpha}$; $N \simeq \theta^{-2}\,I$. Therefore, even if $\alpha \geq 1$,
we may re-observe the GRB within diluted time scales. One may easily notice
a remarkable event just few days after the GRB980425: the GRB980430 (trigger
6715). It is the fourth GRB after SN1998bw and its location is within $\sim
4 \sigma$ from the April event direction (whose error angle is $3.5^\circ$).
The Poisson probability to occur by chance is not negligible ($\leq 2 \cdot
10^{-2}$). However the recent GRB event GRB980712, Batse Trigger 6917 only
within $1.6 \sigma$ from the angular direction of GRB980425 is also very
possibly ($\leq 3 \cdot 10^{-2}$) a repeater signature of the precessing
gamma jet. The additional association of a GRB trigger 6918 nearly 15 hours
later, with a wider error angle makes this combined probability to occur a
rare chance ($10^{-4} \div 10^{-3}$). The duration time of the intrinsic
time scales of progenitor GRB980425 and secondary repeater GRB980712 are
related to the corresponding amplification on factor defined as in equation
\ref{eq13}\
\begin{equation}  \label{eq20}
\frac{\Delta \tau_{04}}{\Delta \tau_{07}} \simeq \frac{20 \,s}{4\,s} = 5
\simeq \frac{a_2}{a_1}
\end{equation}
\newline
where $a_2 \simeq 500$ and now we derive $a_3 \simeq 100$. This value offers
an indication of the gamma jet intensity evolution. Indeed the peak
luminosity flux scales as:\newline
\begin{equation}  \label{eq21}
\frac{L_{04\,\gamma}}{L_{07\,\gamma}} \simeq \frac{I_2\,\theta_2^{-3}}{%
I_3\,\theta_3^{-3}} \simeq \left( \frac{t_3}{t_2} \right)^{-\alpha} \,\left(
\frac{a_2}{a_3} \right)^{\,3} \leq 3.5
\end{equation}
\newline
where $t_3 \sim 78 \; day \sim 7 \cdot 10^6\,s$ while $t_2 \sim 2 \cdot 10^5
\, s$. The total fluence is \newline
\begin{equation}  \label{eq22}
\frac{N_{04}}{N_{07}} \simeq \frac{<L_{04\,\gamma}>}{<L_{07\,\gamma}>} \,%
\frac{\Delta \tau_{04}}{\Delta \tau_{07}} \simeq \left( \frac{t_3}{t_2}
\right)^{-\alpha} \,\left( \frac{a_2}{a_3} \right)^2 \,\frac{\Delta \tau_{04}%
}{\Delta \tau_{07}} \geq 3 \;\;\; .
\end{equation}
\newline
These values are at least comparable with the observed ones and offer a
first suggestive probe of the GRB980425 repeater nature and the imprint of a
precessing gamma jet blazing at least twice to us. The repeater nature of
GRB980425 implies a clearer link between GRB and SGRs. The latter are old
decayed precessing gamma jets observable in their SNR regions only within
our extended galactic halo, and possibly at lower fluxes, from nearby
galaxies within local group.

\section{The SGR-GRB link, the new SGR1627-41 and the multi precessing gamma
jet (GJ).}

The discover of four identified SGRs in the last 20 years: SGR0526-66,
located in Large Magellanic Cloud, and three galactic sources SGR1806-20,
SGR1900+14 and the recent discovered SGR1627-41 toward galactic centre offer
a deeper understanding of a precessing gamma model. The old jet stages,
after a first SN event, as a precessing gamma jet, is possibly feeded by an
accretion disk and/or a companion star (white dwarfs, NS ...) whose presence
modulate the gamma jet directions in a precessing (quasi periodic)
processes. In the most naive approximation the angular size between the jet
and the source-Earth axis, $\theta_1$ (equations 8-11), is evolving by the
binary angular velocity $\omega_b$ as $\theta_1(t) = \sqrt{\theta_{1 m}^2 +
(\omega_b t )^2}$; [10]. The time t=0 corresponds to the maximal intensity
at minimal impact angle $\theta_{1 m}$. The GJ differential fluxes
(equations 8-9-10) would be, in this case, very smooth and periodic. However
the pulsar (or BH) source of the Jet must reflect its spin ($\omega_{psr}$)
frequency in angle $\theta_1(t)$ evolution if his angular momentum axis is
not coincident with the GJ axis. This fast spinning will, usually, inprint
the ``trembling'' millisecond behaviour of most structured GRBs. Finally the
possible anisotropy of the GJ object (related for instance to its own
different inertial momentum, orthogonal and parallel, to the spin axis $%
I_\perp$, $I_{||}$) would modulate by nutation the beam-observer angle $%
\theta_1$ by an angular velocity $\omega_N \sim \omega_{psr}\, \frac{I_\perp
- I_{||}}{I_{||}}$. The combined multi-precessing and spinning beam angle
will describe in the sky a multiple cycloidal (or epicycloidal) trajectory
(almost stochastic) described (in present approximation) by \newline
\begin{equation}  \label{eq25}
\begin{array}{c}
\theta_1(t) = \sqrt{[\theta_{1 m} +\theta_{psr} \cos (\omega_{psr} t +
\phi_{psr}) + \theta_N \cos (\omega_{N} t + \phi_{N})]^2 +} \\
\\
\overline{+\;\;[\omega_{b} t + \theta_{psr} \sin (\omega_{psr} t +
\phi_{psr}) + \theta_N \sin (\omega_{N} t + \phi_{N})]^2} \;\;.
\end{array}
\end{equation}
\newline
These 4 amplitude angle parameters and their 3 arbitrary phases offer a wide
arsenal to mimic most GRBs morphology and signature.\newline
For the SGRs the event is, usually,less structured than GRBs and it simply
implies a smaller (or null) angle $\theta_{psr}$ between jet and angular
momentum directions. Otherwise $\theta_{psr} \gg \theta_b ,\,\theta_N$.
However the crossing of a precessing gamma beam toward observer is
observable (because of the nearer locations of the sources) even at wider
jet cone envelopes. This implies, as observed, softer spectra for SGRs and
less structured one. A first rough estimate of the beaming solid angle is
offered by the ratio $\frac{\Delta \tau_{obs}}{\Delta \tau_{SGR}} \approx
\frac{10\,yr}{1\,s} \sim 3 \cdot 10^8 \simeq \theta^{-2}$, in agreement with
assumed Lorentz factors $\gamma_e \simeq 10^4$. In particular the jet
signature would be observable for nearest SGRs. Indeed the last SGR1627-41
is found in a supernova remnant G337.0-41 near galactic central regions. Its
radio (843 Mhz) plerion image suggests the presence, in between the two
radio lobes, of a jet source (of the plerions). Indeed the SGR1627-01
location (R.A. $247^\circ$, Dec $\sim\, 47^\circ\,33^\prime$) lays between
the two lobes but above them. What is a possible reason of the slight
asymmetry of the bent jet?\newline
We already proposed long ago as evidence for precessing gamma jets, the
spectacular [12] image of the twin rings around SN1987A. We understood their
puzzling existence as the spraying of a conical precessing jets on spherical
remnant of the red giant progenitor: we suggested that dipolar interaction
of the jet object with magnetic field of the binary companion, at nearest
perihelion, is responsible for large bending of the jets and consequent
slight asymmetry of the two twin rings.\newline
Here in analogy, we do understand the asymmetry on the SGR1627-41 location
as due to the GJ binary interactions. In a rough approximation one may
imagine the jets as ``rowing'' at any companion encounter and propelling the
NS jet in opposite plerion directions [13]. This may also explain the
predictable high velocity needed to push the SGR1627-41 far away from its
original birth place just in between the plerion SNR centre.\newline

\section{Conclusions.}

We questioned on the huge ratio between observational probability,
luminosity, energy for GRB events far away or nearby as GRB971214 versus
GRB980425. The six-seven order of magnitude ratios is against any ideal
fireball/hypernova candle model. Some authors are still dubious on GRB
nature of near events. New populations of fine tuned GRBs has been already
proposed [6]. Here we have shown the general credibility of GRB980425 -
SN1998bw connection, greatly more reliable than any cosmic associations
(equation 4). We proposed to solve the multiples GRBs puzzles within an
unique GRB model: a precessing gamma jet.\newline
Its different geometrical observational features may solve, at once,
opposite puzzles: the mysterious and unexplained rarity of near GRB (April)
event (seven orders of magnitude in equation 7) compensated by its apparent
low peak gamma luminosity (seven orders of magnitude in equation 3); the
nearby SN1998bw has been observed at the wider jet cone periphery
(increasing the observational probability) but its gamma flux is (in
comparison with better on-line GRB981214 jet event) more weak and diluted
and softer event. In a sentence, near GRB980425 pays its extreme statistical
rarity by blazing us just out of the beam, with an extremely low gamma flux.
GRBs and SGRs are within an unique precessing gamma jet model observed at
different beam-angle and at different ages. Extreme powerful beamed cosmic
GRBs are hidden at highest redshift by dust and luminosity dilution. While
at birth (near their isotropic SN optical event) they are observable if at
cosmic distances within the GJ inner beam, at older ages their gamma
intensities are decayed and the GJ are lost in the background noise. On the
contrary nearby and/or young GJ may blaze us twice or more. This might be
already occurred for recent GRB980425. Indeed GRB980430 Batse trigger 6715
is found within $4\,\sigma $ error angle from SN1998bw five days later and
in particular GRB980712 (trigger 6917 and, possibly,  6918) is located
within $1.6\,\sigma $ error angle from SN1998bw. These correlated events are
favoring a GJ model over any one shoot fireball/hypernova S-CGR ``candle''
models. Our present solution of a unique GRB-SGR model is able to satisfy at
once the rare probability (from Eq.7 to Eq. 17) puzzle as well as the
far/near gamma luminosity puzzle (from Eq.2-3 to Eq.18-19) and the duration
times scales (equation 20). \newline
The present dynamic model similar to earlier stationary cosmic [15] and
galactic [8-14] models prescribes repeatability [16] of nearest GJ and their
non-thermal equilibrium. Indeed predicted intensity evolutions seem to
satisfy observations (equations 22,23,24). Spectacular evidences of the
precessing jet are found in the recent SGR1627-41 radio plerions, originated
by its SGR jets; we proposed also optical evidences [12][13], related to
SN1987A rings, as recently noticed [13-14-17-7]. We foresee the presence of
a runaway pulsar relic in south-east direction respect to the original
SN1987A centre (in opposite direction respect to intuitive expectation) and
an optical jet source at SGR1627-41 centre. We might expect SN1998bw
bursting again in gamma within a year from now. Finally the detailed images
of nebula NGC6543 shown by Hubble (``Cat Eye'' nebula) and its thin luminous
jets fingers , the exceptional and inexplicable double cone sections found
in Egg Nebula CRL2688 are probably the most detailed view showing an active
precessing GJ in space seen on a lateral side.\newline

\section*{Acknowledgments}

The author wishes to thank Prof. G. Salvini for kind support, Drs R.
Conversano, F. Chiarello, A. Salis , A. Aiello for useful conversations and
help.

\end{document}